\documentstyle[preprint,aps]{revtex}

\draft

\begin{document}
\preprint{NCL94-TP5}

\title{The Canonical Partition Function For Quons}

\author{J. W. Goodison\cite{email,www} and D. J. Toms\cite{emailtwo} }
\address{Physics Department, University of Newcastle upon Tyne,\\
Newcastle upon Tyne, NE1 7RU, U.K.}
\date{4th October 1994}

\maketitle

\begin{abstract}
  We calculate the canonical partition function $Z_N$ for a system
of $N$ free
particles obeying so-called `quon' statistics where $q$ is real
and satisfies $|q|<1$ by using simple counting arguments. We observe
that this
system is afflicted
by the Gibbs paradox and that $Z_N$ is independent of $q$. We
demonstrate that
such a system of
particles obeys the ideal gas law and that the internal energy $U$ (
and hence
the specific heat
capacity $C_V$ ) is identical to that of a system of $N$ free
particles obeying
Maxwell-Boltzmann statistics.
\vskip 0.5in
\center{To be published in {\bf Physics Letters A} }
\vskip 0.5in
\end{abstract}


\section{Introduction}

Since first introduced by Kulish and Rashetikhin \cite{kulras} in 1981
within
the framework of an
integrable model in quantum field theory, quantum groups
\cite{jim1,jim2,jim3,wor1,wor2} have become
a topic of great interest to theoretical physicists with many
suggested
applications ( e.g. see
\cite{passal,wit,ger,kausal,iwa,solkat}~). The simplest
of these structures, ${\rm SU}_q (2)$, is easily constructed from a
pair of
commuting operators
$a_1$ and $a_2$ plus their hermitian adjoints which satisfy the
$q$-boson
algebra ( first introduced
by Biedenharn \cite {bie}, Macfarlane \cite{mac} )

\begin{equation}
a_i a_i^{\dagger} - q^{  1 / 2  } a_i^{\dagger} a_i = q^{ - N_i/2 }
 \quad \quad i=1,2
\end{equation}
  where $q$ is a real deforming parameter and $N_i$ is the number
operator of
the $i$'th mode.~( We
note that the case $q=-1$ is not well defined since the commutator of
the step
operators $J_{+}^q$
and $J_{-}^q$ of ${\rm SU}_q (2)$ is also badly defined ). It is
obvious that as
$q \rightarrow 1$
the relation (1) goes over into the canonical commutation relation
appropriate
for the description
of bosons.

  The relation (1) exists in many equivalent forms; by making the
transformation

\begin{equation}
b_i =  q^ { N_i/4 } a_i
\end{equation}
  one arrives at the single-particle form of Greenberg's `quon'
algebra
\cite{gre1} viz.

\begin{equation}
b_i b_i^{\dagger} - q b_i^{\dagger} b_i = 1
\end{equation}
  where different modes commute ( a consequence of the commutation of
the
$a_i$'s ). We have now
cast (1) into a form which clearly interpolates between bosons and
fermions and
is perhaps less
cumbersome due to the removal of the number operator factors on the
RHS. For the
remainder of this work, we will consider the $b_i$'s to be fundamental
objects,
not composites of the $a_i$ and $N_i$ operators. It should also be
noted that
equation (3) originally arose in the work of Arik and Coon
\cite{arikoo}.

  The purpose of this paper will be to investigate the statistical
thermodynamics of a system of $N$ free ( i.e. the single
particle Hamiltonian is assumed to be proportional to the number
operator )
`quons' obeying the single-particle relation (3).
However, as we shall see in section
II,
this system is purely bosonic ( complete with Bose-Einstein
condensation ) with
the effects of the
deforming parameter only appearing in correlation functions {\sl if
the $b_i$'s
are assumed to
commute }. We are therefore motivated to
depart slightly from (3) to Greenberg's multi-mode `quon' algebra (
also studied
in \cite{moh1} and in an `anyonic' form  in \cite{solkat2}~)

\begin{equation}
b_i b_j^{\dagger} - q b_j^{\dagger} b_i = \delta_{ij}
\end{equation}
  i.e. different modes will be assumed to $q$-mute rather that
commute. ( This is, of course, a different physical system to that
where the
$b_i$'s are assumed to commute ).

  We assume the existence of a vacuum state $|0>$ which satisfies

\begin{equation}
b_i |0 > = 0
\end{equation}
  and build up the Fock space by applying monomials in creation
operators to the
vacuum state. We define the $n$-particle state
$|j_1j_2j_3.....j_n>$ by

\begin{equation}
b_{j_1}^{\dagger}b_{j_2}^{\dagger}b_{j_3}^{\dagger}.....b_{j_n}
^{\dagger} |0>
=|j_1j_2j_3.....j_n>
\end{equation}
  for all values of $q$. Fivel \cite{fiv} and Zagier \cite{zag} showed
that the
resulting Hilbert space is positive
definite if $|q| \le 1$ - for larger values of $|q|$ states with
negative
squared norm arise and the
probability interpretation of quantum mechanics is lost. We therefore
will
restrict ourselves to
$|q|<1$. We note that in this case, the $n$-particle state has no
definite
symmetry properties under
particle interchange - this is a consequence of the simple observation
that the
relation

\begin{equation}
b_i b_j = q b_j b_i
\end{equation}
  can not be consistently imposed unless $q \pm 1$. ( This is easily
seen upon
interchanging the indices in (7). )

  The structure of this paper is as follows. In section two, we derive
the
canonical partition
function $Z_N$ for $N$ free particles quantized according to the
relation (4).
We show that $Z_N$ is
independent of $q$ and is therefore equal to the expression given by
Greenberg
\cite{gre2} for the
special case $q=0$ ( ``infinite statistics'' ). In section three, we
examine
some of the
thermodynamic
properties of this system and observe that it is afflicted by Gibbs
paradox. We
also observe this
effect in the grand canonical formalism. Concluding remarks are given
in section
four.

\section{The Canonical Partition Function}

  We begin the derivation by reviewing Greenberg's calculation of the
canonical
partition function for $N$ free particles obeying ``infinite
statistics'' i.e.
they
obey the relation

\begin{equation}
b_i b_j^{\dagger} = \delta_{ij}
\end{equation}
  which is clearly a special case of the quonic algebra (4). An
advantage of
studying this system in detail is that it is generally much simpler
than general
$|q|<1$, although often having the same qualitative results, as was
again noted
by Greenberg. We observe that the $n$-particle states are orthogonal
since it is
clear that

\begin{equation}
< j_1 j_2 \dots j_n | k_1 k_2 \dots k_n > = \delta_{j_1 k_1}
\delta_{j_2 k_2 }
\dots \delta_{j_n k_n}
\end{equation}
  where the last equation follows directly from the defining relation
(8). The
computation of $Z_N$ is now possible; we proceed as in the
Maxwell-Boltzmann
case ( for example, see
\cite{pat} ). We assume that we are given a set of occupation numbers
$\{ n_i
\}$ which satisfy

\begin{equation}
N = \sum_i n_i
\end{equation}
  It is then simple to see that the number of orthogonal quantum
states $g \, (
\{ n_i
\} )$ is given by

\begin{equation}
g \, ( \{ n_i \} ) = { N! \over \prod_i ( n_i !) }
\end{equation}

  We now relate the partition function $Z_N$ to the single particle
partition
function $Z_1$ - however $Z_1$ is unchanged from the normal bosonic
case and is
therefore given by the expression

\begin{equation}
Z_1 = { V \over \lambda^3 }
\end{equation}
  which is easily obtained by standard methods, and we have introduced
the mean
thermal wavelength of the particles $\lambda$, where $\lambda$ is
given by

\begin{equation}
\lambda = \biggl( { 2 \pi \over m T } \biggr)^{1 \over 2}.
\end{equation}
  ( We use units throughout such that $\hbar=k_B=1$ ). The calculation
is
completed by noting that the number of orthogonal quantum states is
the result
obtained for the Maxwell-Boltzmann distribution except for the Gibbs
${N!}^{-1}$
factor; hence $Z_N$ for particles quantized according to (8) is

\begin{equation}
Z_N = \biggl( { V \over \lambda^3 } \biggr)^N
\end{equation}

  We will now reason that $Z_N$ is $q$-independent for $|q|<1$ - this
was proved
recently by
Werner \cite{wer}, although we shall obtain this result through simple
counting
arguments. The
critical observation is that all of the vectors in our positive
definite Hilbert
space are linearly
independent, and are orthogonal for the special case $q=0$. However
for general
$|q|<1$, we may
imagine a change of basis to an orthogonal set of states
$|{j_1}^{'}{j_2}^{'}{j_3}^{'}.....{j_n}^{'}>_o$ so that

\begin{equation}
|j_1j_2j_3.....j_n> = \sum_ {
{j_1}^{'}{j_2}^{'}{j_3}^{'}.....{j_n}^{'} }
c_{ {j_1}^{'}{j_2}^{'}{j_3}^{'}.....{j_n}^{'} }
|{j_1}^{'}{j_2}^{'}{j_3}^{'}.....{j_n}^{'}>_o
\end{equation}
  This is many ways resembles Greenberg's use of $q=0$ operators as
building
blocks for
general $|q|<1$ - see section V of \cite{gre1} for a complete
discussion.

  Now we have a similar problem to that at the start of this section
in that we
have a
collection of orthogonal $n$-particle states. With the aid of the
remark that we
will
require $p$ orthogonal vectors to span the same vector space defined
by $p$
linearly
independent vectors, it should become clear that $ g \, ( \{ n_i \} )$
is given
by (11) for
all $|q|<1$. This in turn ensures that $Z_N$ is independent of $q$
within this
range.

  We also hope it is now apparent why the alternative physical system
described
with commuting $b_i$'s is trivial. It
is clear from the above discussion that the canonical partition
function of a
free system is
determined entirely by the number of orthogonal quantum states - if
this is
unchanged
from the standard Bose case then the resulting $Z_N$ will also be
unchanged from
the
Bose case. Hence, condensation will occur {\sl at the same temperature
as in the
undeformed i.e. $q=1$ case}. The only effect of the deformation
parameter will
be in
the correlation functions \cite{vokzac}, for example

\begin{equation}
< b_i^{\dagger} b_i > = { 1 \over {\rm e}^{ \beta \omega_i } - q }
\end{equation}
  where $< \dots >$ indicates the thermal average, and $\omega_i$ is
the
constant of proportionality
in the single particle Hamiltonian. ( This result, along with other
thermal
averages, is derived in the appendix ).

\section{ Quon Statistical Mechanics }

  Armed with the canonical partition function as found in section II,
we may now
investigate some
of
the physical properties of the free `quon' gas. The starting point
will the
Helmholtz free energy,
$A$, of the system. This is easily expressed as ( where we have used
(13) )

\begin{equation}
A = N T \bigl( 3 {\rm ln} \lambda - {\rm ln} V \bigr)
\end{equation}

  Immediately from the above expression we obtain relations for the
pressure
$P$, entropy $S$ and
the chemical potential $\mu$ via

\begin{equation}
P = - \biggl( { \partial A \over \partial V } \biggr) = { N T \over V
}
\end{equation}

\begin{equation}
S = - \biggl( { \partial A \over \partial T } \biggr) = { 3 \over 2 }
N  -
{
A \over T }
\end{equation}

\begin{equation}
\mu = \biggl( { \partial A \over \partial N } \biggr) = { A \over N }
\end{equation}

  It is perhaps worth making comments about all three of the results.
Firstly,
it is interesting to
observe that the free `quon' gas obeys the ideal gas law. Secondly, we
note that
as $T \rightarrow
0 \; ( + \infty) \;$, the entropy $S \rightarrow - \infty \; ( +
\infty )$.
Finally, we note that
there exists a temperature $T_c$ where the
chemical potential vanishes, where $T_c$ is given by

\begin{equation}
T_c = { 2 \pi \over m V^{2 \over 3} }
\end{equation}

  Using the quantities we have derived above, we can now also
construct
expressions for the total
energy of the system, $U$, and the specific heat capacity per unit
volume $C_V$
via

\begin{equation}
U = A + TS = { 3 \over 2 } N T
\end{equation}

\begin{equation}
C_V = \biggl( { \partial U \over \partial T } \biggr) = { 3 \over 2 }
N
\end{equation}
  We note that the results (22) and (23) are identical to those
obtained for a
Maxwell-Boltzmann
gas consisting of $N$ free particles.

  It is necessary to view all of these results with a cautious eye,
however.
Consider for example
the Helmholtz free energy of the system $A$ for a fixed density $\rho
= N/V$. By
definition $A$
must be an extensive quantity i.e. proportional to the size of the
system;
substituting into (17) we
observe that $A$ contains a term which grows like $ N \,{\rm ln}\, N
$. This is
a well known
problem in statistical mechanics and is known as Gibbs paradox - it is
for
precisely this reason
that the Gibbs ${N!}^{-1}$ correction term was introduced. However, in
our case
we can not
introduce an {\sl ad hoc} correction term; the weightings have been
determined
by quantum
statistical mechanics.
  As yet, we have only considered the canonical system. As may be
anticipated
though, Gibbs paradox
is also evident in the grand canonical formalism. The grand canonical
partition
function, ${\cal
Z}$, is defined as

\begin{equation}
{\cal Z}  = \sum_{N=0}^{\infty} z^N Z_N
\end{equation}
where $z = \exp ( \beta \mu )$ is the fugacity. The average particle
number,
$<N>$, is then given by

\begin{eqnarray}
<N> && = z { \partial ( \ln {\cal Z} ) \over \partial z } \\
&&= { zV \over \lambda^3 - zV }
\end{eqnarray}
  where we have used our earlier result for $Z_N$. Here the problem is
that
$<N>$ should be an extensive variable of the system; however it is
manifestly not proportional to the volume $V$. We also note that as $T
\rightarrow T_c$, the
average
particle number $<N>$ diverges since $z \rightarrow 1$ and $ \lambda^3
\rightarrow V$. ( This is an
equivalent statement to Werners assertion that there exists a volume V
at a
given temperature for
which $<N>$ diverges ).

\section{Discussion}

  By considering the number of orthogonal quantum states $ g \, ( \{
n_i \} )$
for a given set of
occupation numbers $ \{ n_i \} $, we have calculated the canonical
partition
function for $N$ free
`quons' and shown it to be independent of $q$ for $|q|<1$. One may
then use this
result to show that
this system shares many physical characteristics with a system of $N$
free
particles obeying
Maxwell-Boltzmann statistics; specifically the ideal gas law holds,
and the
total energy of the
systems is the same.

  We also observe that there exists a ( volume dependent ) critical
temperature
$T_c$ at which the
chemical potential vanishes. It is found that in the grand canonical
formalism
the average particle
number $<N>$ diverges at this critical temperature. Whether or not
this is some
kind of condensation
phenomena appears to be an open question.

  More important perhaps is attempting to determine whether this
system is
`physical'. The
appearance of Gibbs paradox due to the form of the function $ g \, (
\{ n_i \}
)$ and all the
problems associated with the mixing of gases is a severe blow. As
Werner notes,
the thermodynamic
limit just does not make sense for this system.

  There appear to be two possible routes out of this troublesome
situation.
Firstly, one may claim
that due to very large interaction times, the system has not yet
reached an
equilibrium state, as
was suggested by Greenberg and Mohapatra \cite{gremoh}; this would
then enable
cosmological bounds
to be placed on $q$.

  The second alternative is perhaps more palatable. As was noted
within the
introduction, it is not
possible to relate $b_i b_j$ to $b_j b_i$ in a consistent way for all
$i$ and
$j$. However, it is
possible to impose

\begin{equation}
b_i b_j = f(q) b_j b_i \hskip 1in i>j
\end{equation}

\begin{equation}
b_i b_j = { 1 \over f(q) } b_j b_i \hskip 1in i<j
\end{equation}
  where an obvious candidate is $f(q)=q$ although other choices are
possible.
Effectively, this
alters the number of orthogonal quantum states, thus leading us away
from Gibbs
paradox. We hope to
study this possibility in a future work.

\acknowledgements

   JWG would like to thank the E.P.S.R.C. for financial support. DJT
would like
to thank the
Nuffield Foundation for their support.

\appendix
\section*{Calculation of thermal averages}

   In section III, we quoted the result for the thermal average of the
quantity
$b_i^{\dagger} b_i$. For the sake of completeness we derive that
expression
here.

  Since we know both the form of the one-particle energy spectrum (
due to us
having chosen the free Hamiltonian ) and that our single-particle Fock
space
consists of states
$|0>, |1>, |2>, \dots $ we may compute the thermal average of any
operator $\hat
O $ using

\begin{equation}
< \hat O > = { \sum_{n=0}^{\infty} < n | \hat O {\rm e}^{ - \beta H}
|n >
\over \sum_{n=0}^{\infty} < n | {\rm e}^{ - \beta H} |n > }
\end{equation}

  The simplest thermal average to compute ( and perhaps the biggest
source of
confusion~) is that of the number operator $N$. Using the above
definition, we
find that

\begin{equation}
< N > = { \sum_{n=0}^{\infty} n {\rm e}^{ - \beta n \omega}
\over \sum_{n=0}^{\infty} {\rm e}^{ - \beta n \omega } } = { 1 \over
{\rm e}^{
\beta \omega } - 1 }
\end{equation}
  i.e. a result which is {\bf unchanged from the bose case}. This
implies that (
as correctly stated by Vokos and Zachos \cite{vokzac}~) $q$-deformed
black-body
spectra ( e.g. \cite{bab}~) are mythical. However, it does not imply
that `quon'
thermodynamics is the same as bose thermodynamics since the definite
symmetry
under particle underchange of the bose case has been removed.

  We now compute the thermal average given in section III. Using the
defining
relation (3), it is easy to show that

\begin{equation}
b_i^{\dagger} b_i |n> = [n]_q |n> \hskip 0.6in b_i b_i^{\dagger} |n> =
[n+1]_q
|n>
\end{equation}
  where the square bracket function $[x]_q$ is given by

\begin{equation}
[x]_q = { x^q - 1 \over x - 1 }
\end{equation}
  Using this, it is a matter of simple arithmetic to obtain

\begin{equation}
< b_i^{\dagger} b_i > = { 1 \over {\rm e}^{ \beta \omega_i } - q }
\hskip 0.6in
< b_i b_i^{\dagger} > = { {\rm e}^{ \beta \omega} \over {\rm e}^{
\beta \omega_i
} - q }
\end{equation}

  We observe from (A5) that the ratio of the correlation functions is
unchanged
from the bose case. However, in $q$-deformed field theories ( since
the emission
and absorption rates are proportional to $[n+1]_q$ and $[n]_q$
respectively )
the emission rates are changed - see \cite{vokzac} for a full
discussion.

  One should also be aware of the perils of interpreting the quatity
$<b_i^{\dagger}b_i>$ as the average number of particles in the $i'th$
energy
level \cite{gesu}. This is ( obviously ) only correct when the number
operator
is given by $b_i^{\dagger}b_i$ i.e. when $q = \pm 1$.

\end{document}